\documentstyle[12pt]{article}
\begin{document}
\pagestyle{empty}
$\ $


\vskip 1.5 truecm

\centerline{\bf
An Attempt to Construct the Standard Model with Monopoles
}

\

\centerline
{Tanmay Vachaspati}
\smallskip
\centerline
{\it
Physics Department, Case Western Reserve University
}
\centerline
{\it
Cleveland, OH 44106.
}

\vskip 1. truecm

\begin{abstract}
We construct a model in which stable magnetic monopoles
have magnetic charges that are identical to the electric
charges on leptons and quarks and the colored monopoles are
confined by strings in color singlets.
\end{abstract}
\clearpage
\pagestyle{plain}

The similarities between magnetic monopoles connected by strings
and quarks connected by chromoelectric flux tubes has been the
basis for speculation over the last few decades that there may
be a direct correspondence between the two \cite{yn}. This
speculation is
further fueled by the recent developments in supersymmetric
theories where duality transformations can be found that
relate the spectrum of particles in one theory to
the spectrum of monopoles (and vice versa) in another dual
theory \cite{asen}.
In spite of the similarities in the pictures of quarks and
monopoles, a specific model in which
monopoles are confined in composites just as quarks
are confined in baryons and mesons is lacking in the
literature. It is the purpose of this paper to construct
such a model. As a bonus, the model is found to contain
other monopoles which do not get confined
and which have a charge
spectrum similar to that of the standard model leptons.
The successful replication of the charge spectrum of the
fermionic sector of the
standard model by monopoles seems quite miraculous and
leads us to speculate that perhaps the real world fermions are
indeed monopoles of some grand unified bosonic theory.
The difficulties likely to be encountered in developing
such a scenario are discussed towards the end of the paper.

We start by listing the desirable features of the model that
we are looking for. These are:

(i) The model should contain magnetic monopoles that are confined
in twos and threes. This feature might be thought of as
``color'' confinement and so we might want the monopoles to
carry $SU(3)$ charge.

(ii) The monopoles should also carry an $SU(2)$ charge which may
be thought of as ``weak'' charge.

(iii) A monopole ``cluster'' (confined monopoles) should have the
ability to carry net magnetic charge which may be identified with
$U(1)$ ``electromagnetic'' charge.

A model that seems to satisfy all these requirements is one in which
the (continuous) symmetry breaking pattern is:
\begin{equation}
\tilde{SU} (5) \rightarrow \tilde U(1)
\label{model}
\end{equation}
where the final $\tilde U(1)$ is to be thought of as the
dual electromagnetic symmetry group. One way to analyze the
monopole and string content \cite{preskill}
of this symmetry breaking would be to construct all the
embedded string solutions \cite{tvmb}
and the ``incarnations'' of topological monopoles \cite{tktv}.
Another way, which is simpler, is to imagine that
the symmetry breaking in (\ref{model}) occurs in stages. For
example,
\begin{equation}
\tilde {SU}(5) \rightarrow
[\tilde{SU}(3)\times \tilde{SU}(2)\times \tilde{U}(1)']/Z_6 \\
\rightarrow [\tilde{SU}(3)\times \tilde{U}(1)]/Z_3 \\
\rightarrow \tilde U (1)
\label{stages}
\end{equation}
The first step can be achieved if an $\tilde{SU}(5)$ adjoint Higgs
($\Phi_{24}$) gets a vacuum expectation value (vev).
The second symmetry breaking occurs
by the vev of an $\tilde{SU}(2)$ fundamental (which can arise
from an $\tilde{SU}(5)$  fundamental).
The third symmetry breaking occurs if three $\tilde{SU}(3)$ adjoints but
$\tilde U (1)$ singlets get vevs since it is known
that the (generic) vevs of $N$ adjoints of $SU(N)$
break the symmetry down to $Z_N$.

The reader would have surely noticed that the first two stages
in (\ref{stages}) are
nothing but the symmetry breaking pattern of minimal $SU(5)$ Grand
Unification. Indeed this is true, and luckily for us, the monopoles
in the model
have been studied in great detail \cite{dt} together with their
stability properties \cite{gh, coleman}.

The potential needed for the first symmetry breaking is:
$$
V(\Phi_{24}) = - {{\mu^2} \over 2} {\rm Tr}[\Phi_{24}^2]
                +{ a \over 4} ({\rm Tr}[\Phi_{24} ^2])^2
                +{ b \over 2} {\rm Tr}[\Phi_{24} ^4]
$$
with the constraints: $a > 0 > -7b/15$. This potential is minimized
by the vev:
$$
<\Phi_{24}> = v \ {\rm diag}\left ({1\over 3},{1\over 3},{1\over 3},
              -{1\over 2},-{1\over 2}\right ) \ .
$$
which is annihilated by the generators of $\tilde{SU}(3)$ acting
on the upper left $3\times 3$ elements, $\tilde{SU}(2)$ acting
on the lower right $2\times 2$ elements and by the $\tilde U (1)$
generator given by
$$
Q_1 = {1 \over v} <\Phi_{24}>
= {\rm diag}\left ({1\over 3},{1\over 3},{1\over 3},
              -{1\over 2},-{1\over 2}\right ) \ .
$$
But there are 3 group
elements that are shared between $\tilde{SU}(3)$  and $\tilde{U}(1)'$
which correspond to the center of $\tilde{SU}(3)$  and are:
$$
{\rm exp}\left [i 2\pi n
                 {\rm diag} \left (
 {1\over 3},{1\over 3},-{2\over 3},0,0 \right )\right ]
= {\rm exp} \left [ i 2\pi n Q_1 \right ] , \ \ \ n=0,2,4\ .
$$
There are also 2 group elements that are shared between $\tilde{SU}(2)$
and $\tilde{U}(1)'$  which correspond to the center of $\tilde{SU}(2)$ and
these are:
$$
{\rm exp}\left [i 2 \pi n
                 {\rm diag} \left (
   0,0,0,{1\over 2},-{1\over 2} \right )\right ]
= {\rm exp} \left[ i 2\pi n Q_1 \right ] , \ \ \ n=0,3\ .
$$
So to avoid overcounting the discrete group $Z_3 \times Z_2$ which
is the center of $\tilde{SU}(3)\times \tilde{SU}(2)$,
we must mod out the
unbroken continuous symmetries by $Z_6$.

The monopoles formed at the first symmetry breaking are given by
the first homotopy of the unbroken symmetry group. This means that
we have to construct closed paths on the group manifold that are
incontractible. Each class of homotopically inequivalent paths,
leads to a distinct monopole. Clearly, an incontractible path
is one which wraps around the $\tilde{U}(1)'$  and this can be written as:
\begin{equation}
{\rm exp}[i 6 Q_1 s], \ \ \ s \in [0,2\pi ]\
\label{path}
\end{equation}
where $Q_1$ is the generator of $\tilde{U}(1)'$ and $s$ is a parameter
that labels points on the closed path. But there are other incontractible
paths present too - for example, there is a path that goes
through $\tilde{U}(1) '$ , $\tilde{SU}(2)$ and $\tilde{SU}(3)$ .
This path may be written in the
form of (\ref{path}) but with $6 Q_1$ replaced by
$$
Q_m = Q_3 + Q_2 + Q_1
$$
where, the $\tilde{SU}(3)$  and $\tilde{SU}(2)$ charge operators are
$$
Q_3 = {\rm diag}\left (- {1\over 3},- {1\over 3},{2\over 3},0,0\right )
$$
$$
Q_2 = {\rm diag}\left ( 0,0,0,{1\over 2},-{1\over 2} \right ) \ .
$$
The monopole corresponding to this path is the ``minimal'' monopole and
all other monopoles in the model can be thought of as multiply charged
monopoles of this variety. So the monopoles in the model have
charge $n \times Q_m$, where $n$ is any integer. But the $SU(3)$ charge is
a $Z_3$ charge and the $SU(2)$ charge is a $Z_2$ charge. Hence, the
winding $n$ monopole has magnetic charge:
$$
Q_m^{(n)} = n_3 Q_3 + n_2 Q_2 + n_1 Q_1 \ ,
$$
where,
$n_3 = n ({\rm mod}\ 3)$, $n_2 = n ({\rm mod}\ 2)$ and $n_1 = n$.
The first four columns of
Table 1 display the quantum numbers of monopoles with different
winding numbers.

\begin{table}
\caption{
Quantum numbers on monopoles with winding $n \le 6$ and charges
on standard model fermions in units of the charges on
$(u,d)_L$.
}
\begin{center}
\begin{tabular}{|cccc|cccc|}
\hline
                   \multicolumn{1}{|c}{$n_{~}$}
                 & \multicolumn{1}{c}{$n_3$}
                 & \multicolumn{1}{c}{$n_2$}
                 & \multicolumn{1}{c|}{$n_1$}
                 & \multicolumn{1}{c}{$$}
                 & \multicolumn{1}{c}{$SU(3)_c$}
                 & \multicolumn{1}{c}{$SU(2)_L$}
                 & \multicolumn{1}{c|}{$U(1)_Y$}         \\
\hline
1&1&1&1&$(u,d)_L$    &1 &1  &+1      \\
2&2&0&2&$d_R$        &1 &0  &-2     \\
3&0&1&3&$(\nu ,e)_L$ &0 &1  &-3     \\
4&1&0&4&$u_R$        &1 &0  &+4      \\
5&2&1&5& -           &- &-  &-       \\
6&0&0&6&$e_R$        &0 &0  &-6       \\
\hline
\end{tabular}
\end{center}
\end{table}

\

\

Even though monopoles of arbitrary charge are allowed topologically,
they may not exist for dynamical reasons or may be unstable to
fragmenting into monopoles of smaller winding. Gardner and
Harvey \cite{gh} have argued that monopoles in
the first stage of symmetry breaking are stable only when $n=\pm 1,
\pm 2, \pm 3,\pm 4$, and, $\pm 6$ provided
$\mu_0 < < \mu_8 = \mu_3 /2$ where
$\mu_0$, $\mu_3$ and $\mu_8$ are the masses of the singlet, triplet
and octet components of $\Phi_{24}$.
A crucial observation here is that monopoles with
$n= \pm 5$ are unstable.


In the last four columns of Table 1
we show the $SU(3)_c$, $SU(2)_L$ and $U(1)_Y$
charges on the leptons and quarks of the standard model
in units of the charges on $(u,d)_L$.
A comparison of the left- and right-hand sections of Table 1
suggests the following identifications:
\begin{eqnarray}
(u, d)_L & \rightarrow n=+1 ~~
\nonumber
\\
d_R & \rightarrow n=-2 ~~
\nonumber
\\
(\nu, e)_L & \rightarrow n=-3 ~~
\label{equiv}
\\
u_R & \rightarrow n=+4 ~~
\nonumber
\\
e_R & \rightarrow  n=-6 ~ .
\nonumber
\end{eqnarray}


The monopoles with non-trivial $\tilde {SU}(3)$ and
$\tilde {SU}(2)$ charges are three- and two-fold degenerate
respectively. It is simplest to see this for the monopoles
with non-trivial $\tilde {SU}(2)$ charge but vanishing
$\tilde {SU}(3)$ charge \cite{bais}.
These monopoles exist due to
incontractible closed paths that are entirely in
$[\tilde {SU}(2) \times \tilde U(1)']/Z_2$. Then consider
the two incontractible paths:
$$
g_{\pm} (s) = {\rm exp}\left [
    is\left ( {{{\bf 1}\pm {\bf \tau}_3} \over 2} \right )
                   \right ]
$$
where, $s\in [0,2\pi]$ is the parameter labeling the path.
It can be checked that the path
$g_+$ can be deformed into $g_-$ and so the paths are
topologically equivalent. This would lead one to think that
there is only one monopole. But now consider what happens
when the $[\tilde {SU}(2) \times \tilde U(1)']/Z_2$ gets
broken as in the last stage of (\ref{stages}). Suppose
the generator of the final unbroken $\tilde U(1)$ is
$Q = ({\bf 1}+{\bf \tau}_3 )/2$ as is conventionally taken
in the standard electroweak symmetry breaking. Then the
$g_+$ monopole will continue to have a long range
$\tilde U(1)$ magnetic field but the $g_-$ monopole magnetic
field will get screened. So we should think of the monopoles
with non-trivial $\tilde {SU}(2)$ charge as being
two-fold degenerate with the degeneracy being lifted in
the last stage of (\ref{stages}). Similarly, we should
think of the monopoles with non-trivial $\tilde {SU}(3)$
charge as being three-fold degenerate.


The two-fold degeneracy of monopoles with non-vanishing
$n_2$ is brought out in (\ref{equiv}) since we have to identify
fermion doublets with the $n=+1$ and $n=-3$ monopoles. Similarly,
the three-fold degeneracy of monopoles with non-vanishing $n_3$
means that these monopoles should come in the fundamental
representation of an $SU(3)$.
The reason that we sometimes have to choose to identify the fermions
with antimonopoles ($n<0$) rather than monopoles is so that the
hypercharge assignments tally. (The $\tilde{SU}(2)$ charge on the monopole
is a $Z_2$ charge and so the sign is not important. The $\tilde{SU}(3)$
charge is a $Z_3$ charge and so -2 is the same as +1.)
Remarkably, this identification yields the correct $\tilde {SU}(3)$
charges of the standard model fermions. It also seems
somewhat of a miracle that the $n=\pm 5$ monopoles are unstable
at the same time that we do not observe any fermions of hypercharge
equal to 5/6.

Is the correspondence between the standard model fermions and
stable monopoles some group theory magic? This cannot be entirely
true since the stability of a multiply charged
monopole also depends on the dynamical
requirement that $\mu_0 < < \mu_8 $.
However, it is true that the instability
of the $n=\pm 5$ monopole is independent of the choice of
parameters since it can always fragment into two monopoles
of winding numbers $\pm 2$ and $\pm 3$ \cite{gh}.

The second symmetry breaking in (\ref{stages})
corresponds to the electroweak symmetry
breaking and this is known not to have any topological strings. Hence,
as discussed in Ref. \cite{bais,gh}, the
monopoles carrying $\tilde{SU}(2)$ charge will not get confined by
strings during this symmetry breaking \cite{comment}. The $\tilde{SU}(2)$
gauge fields will get screened and the confined monopole
clusters can only carry long range $\tilde{SU}(3)$  and
$\tilde{U} (1)$  magnetic fields. The electroweak $Z-$string
and Nambu's monopoles \cite{nambu} will also be present in this
symmetry breaking. When dualized, the electroweak monopole will appear
as an electrically charged confined particle.

In the last symmetry breaking stage, the $\tilde{SU}(3)$  factor breaks
down to $Z_3$ which is the center of the group. This symmetry
breaking has non-trivial first homotopy:
$$
\pi_1 (\tilde{SU}(3) /Z_3 ) = Z_3
$$
and so the symmetry breaking yields topological $Z_3$ strings. The
$Z_3$ strings produced at this stage are deformable to the vacuum
if we allow excitations
of the $\tilde{SU}(5)$ degrees of freedom. This means that the $Z_3$ strings
can end on monopoles which carry $\tilde{SU}(3)$  charge. These monopoles
are precisely those that correspond to the quarks ($n=1, -2$ and $4$)
and the quark
monopoles are confined by $Z_3$ strings in chromomagnetic
singlets. But the cluster can still have $\tilde{U} (1)$  charge.

Assuming that this model can be suitably dualized, we would like
to identify the confined monopole clusters with the hadrons and the
unconfined monopoles with the leptons. For example, in this picture,
the proton would be identified with three $n=1$
monopoles that have been
confined with net $\tilde{U} (1)$  charge equal to $+1$. And the unconfined
$n=+3$ monopole would be identified with the left-handed anti-neutrino
and positron doublet. Proton decay might correspond to
the collapse of 3 confined $n=1$ monopoles to form a single
$n=+3$ monopole.

The correspondence between the $\tilde{SU}(5)$ monopoles and the standard
model fermions suggests that, perhaps ``Grand Unification'' should
be based on an $\tilde{SU}(5)$ symmetry group with only a bosonic sector
and the presently observed fermions are really
the monopoles of that theory.
However, as we now discuss, there are numerous challenges that need
to be addressed before this conjecture can be tested.

The first question is that why
should the monopoles be fermions and not bosons?
This problem may already have been resolved due to the discovery
that isopin can lead to spin \cite{rjcr}. The idea is that a
bound state of a charged boson and a monopole forms a dyon that
can have integer or half-integer spin if the {\it iso}spin of the
free boson is integer or half-integer respectively.
Goldhaber has shown that dyons with half-integer spin
also obey Fermi-Dirac statistics \cite{ag}.
These results may be relevant to our construction but there
is also a problem in this approach. The four
(degenerate) dyons that result
from the bound state of a monopole and a charged boson can
have magnetic and electric charges $(\pm 1, \pm 1/2)$ in
suitable units. A duality rotation cannot result in all these
dyons having pure electric charges. In this picture we would
get the standard model fermions as well as light magnetic
monopoles.

Another scheme to convert monopoles to dyons is by the introduction
of a $\theta$ term in the $\tilde{SU}(5)$ action. This term would be
proportional to $\tilde \theta F_{\mu \nu}^a {\tilde F}^{\mu \nu a}$
where $F_{\mu \nu}^a$ and ${\tilde F}^{\mu \nu a}$ are the
$\tilde{SU}(5)$ field strengths and their duals. Note that we have
chosen to denote
the coefficient of the term by $\tilde \theta$ and not by $\theta$
because we are assuming that the model will ultimately be dualized
and that the $\theta$ in the standard model will be different
from $\tilde \theta$. Witten \cite{ew} showed that the presence
of such a term in the action will convert monopoles into dyons
with electric charge
$\tilde e \tilde \theta / 2\pi$ where $\tilde e$ is the
smallest unit of electric charge in the original (undualized)
model.
As the spin of
a dyon is related to the angular momentum in the long range
fields, it seems reasonable to assume that the dyon in the case when
$\tilde \theta = \pi$ should also be treated as a spin $1/2$ object
obeying Fermi-Dirac statistics. The advantage in this approach
seems to be that, since the $\theta$ term is CP violating, there
are only two degenerate dyons present with
magnetic and electric charges $\pm (1,1/2)$.
And now a duality rotation can convert these dyons
into particles carrying only electric charges.

Note that the scheme presented in this paper is based on a
different philosophy than the scheme used in supersymmetric
duality. Were we to start out with a supersymmetric theory,
our model would already contain both bosons and fermions in
supersymmetric multiplets. The monopole solutions would be
in addition to all the supersymmetric particles. In the present
scheme, we have started out with only bosons - hence, the
model is necessarily non-supersymmetric - and the monopoles
are identified with the fermionic sector of another theory -
in this case, the standard model.

An aspect we have not addressed is the three family structure of
the standard model fermions. It is probably possible to get three
families of monopoles by increasing the number of scalar fields
in the model and perhaps introducing new symmetries under the
exchange of these fields \cite{fwaz}. For example, one could consider the
case when the first symmetry breaking occurs by the parallel (but
not necessarily equal) vevs of 3 different adjoints of $\tilde{SU}(5)$.
Indeed, this might be desirable since we also need
three $\tilde{SU}(3)$ adjoints to get vevs during the last stage of
symmetry breaking in (\ref{stages}) and each $\tilde{SU}(3)$
adjoint could come from an $\tilde{SU}(5)$ adjoint. If this does
lead to three families of monopoles - and this is something that
needs to be investigated - it would relate the number of
families to the number of colors in QCD.

A vexing problem is the connection of $\tilde{SU}(2)$ and chirality and
this remains one of the many open questions. Another problem is
to study the stability of the monopoles at strong couplings and
to actually dualize the model. Only when this is done, will it be
possible to determine the masses and interactions of the particles
(dualized monopoles).

Another aspect that we have not addressed is if the picture
can have any cosmological relevance?
Can it be that only bosonic particles filled the universe at some epoch
and, after some phase transitions, got replaced by monopoles which
today look like the fermionic sector of the standard model?

Within the framework of this model, it seems that the bosonic sector
of the standard model will be replaced by a ``dyonic'' sector since,
after the duality transformation, the original electrically charged
particles (Higgses and gauge fields of $\tilde{SU}(5)$)
will get a magnetic description.
A prediction of this picture may be one that we have alluded to
earlier - when the model is dualized, Nambu's electroweak monopoles
\cite{nambu} would appear as electric charges that are confined
by $Z-$electric flux tubes. These would appear as new confined particles.
Also, we expect that the standard model fermions should resemble
solitons (monopoles or dyons) at short distance scales. Then, for example,
head-on collisions of particles at very high energies should
lead to $90^o$ scattering \cite{manton}.

Finally, we should point out that a spectrum of monopoles similar
to the standard model would also result from other symmetry breaking
schemes. This is because the spectrum of stable monopoles
only depends on the first homotopy of
$[\tilde{SU}(3)\times \tilde{SU}(2)\times \tilde{U}(1)']/Z_6$
and any symmetry breaking pattern
starting from a simply connected group and
with this group as an
intermediate symmetry group will at least contain the monopoles
corresponding to the standard model. Different schemes could,
however, contain extra monopoles that would need to be included
in the spectrum. The $\tilde{SU}(5)$ model we have considered
here is the minimal model that contains all the monopoles
that correspond to the standard model fermions.

To conclude, we have found a correspondence between stable monopoles
in an $\tilde{SU}(5)$ theory and the standard model fermions. The
colored monopoles in this model are confined just as quarks
are confined in color singlets. To take the correspondence further
and claim an equivalence between monopoles and standard model
fermions is a much more difficult task. But, if successfully
done, it could shed light on several aspects of the standard
model such as - the charge spectrum of fermions, why the fermions
appear in fundamental representations, the replication of the
fermions in three families, the confinement of quarks and
other issues. To help us out in this daunting task are
the many beautiful ideas that have been proposed over the last
two decades.

\section*{Acknowledgements}

I would like to thank
Ana Ach\'ucarro, Shekar Chivikula, Tom Kephart, Lawrence Krauss,
Glenn Starkman and Frank Wilczek for discussions and comments.
I am especially grateful to John Preskill for his remarks on the
monopole degeneracy. A significant portion of this work was done
at the Aspen Center for Physics and I am grateful for the Center's
hospitality.






\begin{thebibliography}{999}
\bibitem{comment} The fate of monopoles with the ``wrong''
orientation of charge after the second symmetry breaking has
been qualitatively discussed in \cite{bais,gh} but a detailed
understanding is still lacking.
\end{thebibliography}
\end{document}